\documentclass[journal]{IEEEtran}

\ifCLASSINFOpdf
\else

\fi

\usepackage{tabularx}
\usepackage{makecell}
\usepackage{graphicx}
\usepackage{subfig}
\usepackage{caption}
\usepackage[font=footnotesize]{caption}
\usepackage{amsmath}
\usepackage[switch,pagewise]{lineno} % default option is 'left'
\usepackage{lipsum} % filler text
\usepackage{multirow}
\usepackage{rotating}
\usepackage[table]{xcolor}
\begin{document}

\title{Building Trust Profiles in Conditionally Automated Driving}

\author{{Lilit Avetisyan, Jackie Ayoub, X. Jessie Yang, Feng Zhou, \emph{Senior Member, IEEE}}% <-this % stops a space
\thanks{Lilit Avetisyan and Jackie Ayoub contributed equally.}
\thanks{J. Ayoub , L. Avetisyan, and F. Zhou are with the Department of Industrial and Manufacturing, Systems Engineering, The University of Michigan, Dearborn, 4901 Evergreen Rd. Dearborn, MI 48128 USA (e-mail: \{jyayoub, lilita, fezhou\}@umich.edu).}% <-this % stops a space
\thanks{X. J. Yang is with Industrial and Operations Engineering, University of Michigan, Ann Arbor, 500 S State St, Ann Arbor, MI 48109 USA (e-mail: xijyang@umich.edu).}% <-this % stops a space
\thanks{Manuscript received Nov 17, 2022; revised xxx 26, 2022.}}

\markboth{IEEE Transactions on Intelligent Transportation Systems,~Vol.~xx, No.~xx, November~2022}%
{Avetisyan \MakeLowercase{\textit{et al.}}: Building Trust Profiles in Conditionally Automated Driving}

\maketitle

\begin{abstract}
Trust is crucial for ensuring the safety, security, and widespread adoption of automated vehicles (AVs), and if trust is lacking, drivers and the public may not be willing to use them. This research seeks to investigate trust profiles in order to create personalized experiences for drivers in AVs. This technique helps in better understanding drivers' dynamic trust from a persona's perspective. The study was conducted in a driving simulator where participants were requested to take over control from automated driving in three conditions that included a control condition, a false alarm condition, and a miss condition with eight takeover requests (TORs) in different scenarios. Drivers' dispositional trust, initial learned trust, dynamic trust, personality, and emotions were measured. We identified three trust profiles (i.e., \emph{believers}, \emph{oscillators}, and \emph{disbelievers}) using a K-means clustering model. In order to validate this model, we built a multinomial logistic regression model based on SHAP explainer that selected the most important features to predict the trust profiles with an F1-score of 0.90 and accuracy of 0.89. We also discussed how different individual factors influenced trust profiles which helped us understand trust dynamics better from a persona's perspective. Our findings have important implications for designing a personalized in-vehicle trust monitoring and calibrating system to adjust drivers' trust levels in order to improve safety and experience in automated driving.
\end{abstract}

\begin{IEEEkeywords}
Trust profiles, emotion, personality traits automated vehicles.
\end{IEEEkeywords}

% \fi
\IEEEpeerreviewmaketitle

\section{Introduction}

\IEEEPARstart{A}utomated vehicles (AVs) possess the potential to revolutionize the transportation sector by offering safer and more efficient modes of transportation \cite{ayoub2020}. Nevertheless, the widespread acceptance and implementation of AVs heavily rely on users' trust in this technology. In the absence of trust, drivers may exhibit reluctance towards adopting AVs, thereby impeding or even hindering the deployment of this transformative technology. Consequently, it becomes imperative to design AVs in a manner that cultivates an optimal level of trust, encouraging users to accept and embrace them.

Thus, many researchers have conducted studies on trust in automated vehicles (AVs) and other types of automation, with a particular focus on factors that contribute to and influence trust, such as system transparency, reliability, and performance\cite{ayoub2021investigation,luo2022evaluating}. For example, both Ayoub et al. \cite{ayoub2021investigation} and Azevedo-Sa et al. \cite{azevedo2021internal} demonstrated that when an AV exhibited a high level of system reliability and performance, thereby operating as expected without errors, participants were more inclined to trust the vehicle. Additionally, several studies (e.g., \cite{luo2022evaluating, du2021designing, avetisyan2022investigating}) highlighted the importance of enhancing system transparency. By providing information about the system's status and explanations for automation failures, trust in automation and AVs was increased. 

The aforementioned studies have provided valuable insights to aid in the design of automated systems. However, what appears to be lacking and inconsistent is the consideration of the impact of individual differences on trust in AVs, including factors such as age, driving experience, knowledge of automation systems, personality traits, and emotions (e.g., \cite{chen2017effect,ayoub2021modeling, avetisian2022anticipated}), among others. 
%\textcolor{red}{Jessie: I feel the following two sentences do not contradict each other. }
For instance, empirical studies (e.g., \cite{pak2017effect}) indicated that older adults exhibit a higher tendency to overtrust automated systems compared to younger age groups. Conversely, survey findings revealed that older participants (60 years and above: 45.2\%) expressed greater concerns and, therefore, exhibited lower levels of trust in AVs compared to their younger counterparts (18-29 years old: 26.1\%) \cite{schoettle2016motorists}. It is important to note that individuals of different age groups often employ distinct strategies influenced by various factors such as knowledge, personality traits, and driving experience. Moreover, the specific impact of age may vary in different contexts \cite{hoff2015}.

Moreover, the establishment of trust in AVs encompasses both cognitive and emotional factors, with emotions serving as the primary determinant of trusting behavior \cite{lee2004trust}. However, the influence of emotions on trust in AVs has received limited attention from researchers. Although some efforts have been made to explore the role of emotions in trust towards AVs, further investigation is necessary to fully comprehend the depth of this relationship \cite{avetisian2022anticipated,zhang2020effects}. Specifically, additional research is required to gain insight into the intricate nuances of how various emotions impact trust in AVs and how these emotional responses may interact with other factors, such as the reliability of the system, driver personality traits, and prior experiences with AVs.

Another research gap in many previous studies \cite{merritt2008not,szalma2011individual} pertains to the investigation of individual differences, such as dispositional trust and personality, in relation to trust dynamics. However, these studies primarily focused on capturing trust levels before and after experiments or through cross-sectional surveys, providing only a snapshot of trust dynamics. To gain a comprehensive understanding, it is essential to explore how individual factors influence trust dynamics across various profiles. Previous research demonstrated the existence of different trust dynamics, including \emph{oscillators}, \emph{disbelievers}, and Bayesian decision-makers, in the context of human-robot interaction \cite{bhat_jessie2022clusteringTrust}. Therefore, it is imperative to examine the associations between different individual factors and trust dynamics in the domain of human-AV interaction.

Thus, in this study, we explored trust profiles by aggregating various types of individual factors, including dispositional trust, initial learned trust, personality traits, trust-related emotions, and dynamic trust elicited during the experiment. These factors were examined across three levels of system reliability, which were influenced by different types of errors.

First, we employed a user segmentation approach in our user research to assist designers in understanding the behaviors of different trust profiles. This approach is particularly valuable for comprehending user trust in AVs as it enables designers to identify specific factors, particularly those related to individual differences, that influence trust and subsequently design interventions targeting these factors.

Second, building upon previous studies that demonstrated how dispositional trust and initial learned trust encompass numerous individual factors such as age, gender, culture, driving experience, and knowledge in automation \cite{hoff2015}, we minimized the number of factors by investigating dispositional and initial learned trust.

Third, we utilized a data-driven methodology that employed machine learning models. This included using K-means clustering to identify trust profiles, multinomial logistic regression with SHAP (SHapley Additive exPlanations) to validate these profiles \cite{ayoub2022cause,ayoub2022predicting}, and conducting statistical comparisons among the identified trust profiles. By gaining insights into the trust factors associated with different profiles, designers are empowered to develop trust-oriented solutions that effectively address these factors.

Overall, this research contributes to the understanding of trust dynamics in AVs and provides practical guidance for designers to enhance trust through tailored design interventions.

%Recent studies have addressed the social and cultural basis of trust in the human-AV interaction process. 

% With the increasing reports of AV crashes, trust will be a critical factor in safety and AV adoption. Lee and See \cite{lee2004} defined trust as “\emph{the attitude that an agent will help achieve an individual’s goals in a situation characterized by uncertainty and vulnerability}”. Due to the dynamic uncertainty involved in different situations, the changes of trust are dynamic. Thus, it is important to capture trust dynamics over time in order to avoid undesirable trust conditions (e.g., overtrust and undertrust).  

% Researchers have been trying to estimate trust dynamics over time in automated driving \cite{ayoub2021investigation, azevedo2020real} to design trust-aware AV systems that are capable of changing their behaviors based on drivers' trust levels. 
% The reason behind real-time estimation of trust is that people can be trapped in undesirable trust conditions (i.e., overtrust and undertrust) that could lead to inappropriate use of AVs \cite{parasuraman1997humans}. 

% Misuse refers to failures when people rely inappropriately on automation (e.g., using it when it should not be used) \cite{lee2004trust}. Disuse occurs when people reject the capabilities of the AV \cite{lee2004trust}. Thus, by measuring trust dynamics over time, it will be possible to identify and calibrate undesirable trust conditions in automated driving.

\section{Related Work}

In order to understand trust in AVs, it is important to understand the factors that affect trust. Researchers have identified a wide range of factors that impact trust in AVs. Hoff and Bashir (2015) \cite{hoff2015} proposed a taxonomy of these factors based on three types of trust, namely dispositional, learned, and situational trust. In the context of driver-AV interaction, dispositional trust represented drivers' tendency to trust AVs, including typical factors such as age, gender, culture, and personality; learned trust represents the drivers' assessment of the AV based on their past experience (e.g., preexisting knowledge about the AV) or current interaction (e.g., reliability and performance of the AV) with the AV; situational trust is dependent on the interaction between drivers and the automation in specific contexts, including external environment of the interaction between the driver and the AV (e.g., the complexity and risks associated with the task) and internal characteristics of the driver (e.g., the emotional states and cognitive workload of the driver). 

It is crucial to investigate individual factors that influence dispositional trust. For instance, Robert et al. \cite{robert2018personality} found that participants in high-context cultures, such as East Asia (including China, South Korea, Japan) exhibited greater trust in AVs compared to individuals from low-context cultures (e.g., Western Europe and US) when explanations were provided. Personality was also identified to be a vital factor on trust. Chien et al. \cite{chien2016relation} demonstrated that higher levels of agreeableness and conscientiousness in personality traits were associated with increased initial trust in automation, with agreeableness and conscientiousness being two of the dimensions in the Big Five model of personality. However, experimental examination of these factors can often be challenging due to contextualization and potential interactions with other factors, such as age and driving experience. This can result in inconsistent findings, as observed in the effects of age on trust in AVs \cite{pak2017effect, schoettle2016motorists}. Consequently, it is crucial to investigate multiple individual factors concurrently to gain a comprehensive understanding of their overall impact on trust in AVs.

For learned trust, numerous studies examined preexisting knowledge and experience of AVs through surveys and current interaction through experimental studies. For example, Ayoub et al. \cite{ayoub2021modeling} found that knowledge of AVs generally increased trust in AVs while experience in driving decreased trust in AVs. As mentioned previously, during interaction with AV, a high level of system transparency, reliability, and performance generally increased trust in AVs \cite{ayoub2021investigation,luo2022evaluating,azevedo2021internal}. 
However, it is crucial to consider the temporal aspect of trust development, particularly during extended interaction periods with AVs. To explore this, Bhat et al. \cite{bhat_jessie2022clusteringTrust} conducted a study involving sequential decision-making tasks and employed a clustering model to identify distinct trust dynamics and different types of trust profiles.

For situational trust, both external variability and internal factors should be considered to understand trust in AVs. For example, Azevedo-Sa et al. \cite{azevedo2021internal} found that trust in AVs increased over time when internal risk was low (i.e., system reliability was high) while external risks (visibility in driving) did not impact trust significantly. 

While the effects of many other factors on trust were investigated, less is known about the effects of emotions on trust in AVs. Previous research has suggested that positive emotions may foster trust due to their association with feelings of safety and security \cite{myers2016influence}. For instance, Du et al. \cite{du2020examining} discovered that positive emotions improved takeover performance in AVs, subsequently leading to increased trust. Ayoub et al. \cite{ayoub2021modeling} demonstrated a positive correlation between the feeling of excitement and participants' trust in AVs. Conversely, negative emotions have been found to diminish trust, even when unrelated to trust-related decisions \cite{engelmann2019neural}. Myers et al. \cite{myers2016influence} investigated the influence of three negative emotions (anger, guilt, and anxiety) on trust and found that negative emotions with low certainty appraisals (e.g., anxiety) reduced trust, whereas those with high certainty (e.g., anger and guilt) had no discernible effects on trust. Additionally, Ayoub et al. \cite{ayoub2020} revealed that negative emotions such as concerns and worries decreased parents' trust in automated school buses. Thus, it is imperative to examine the effects of emotions and their role in identifying trust profiles in the context of AVs.

\section{Methodology}
\begin{figure*}[h!]
\centering
\includegraphics[width=.9\linewidth]{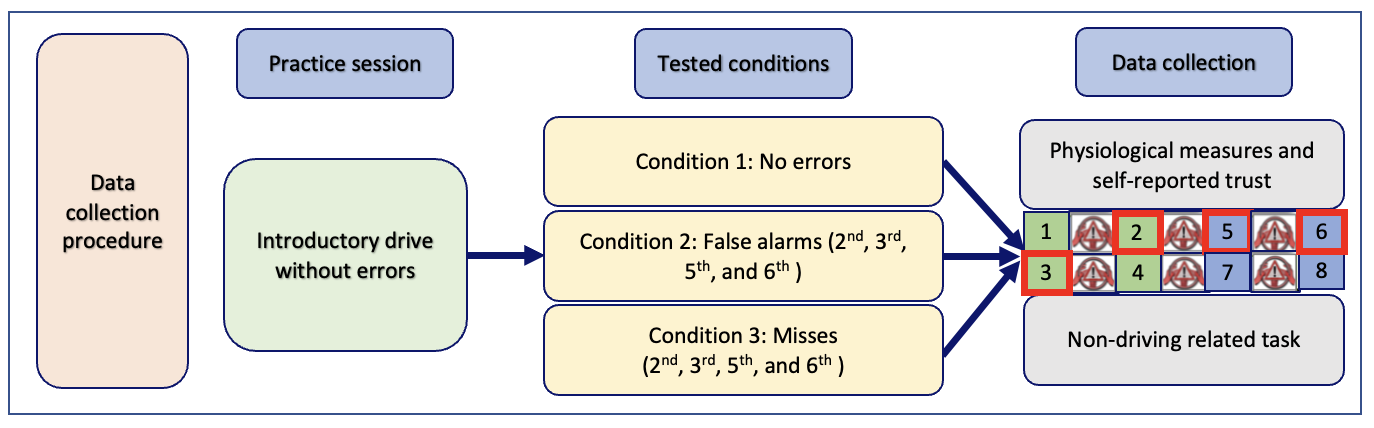}\hfill
\caption{Framework of the proposed study.}
\label{fig:expProc}
\end{figure*}

% in methodology focus on two sections predictions and explanations
Fig. \ref{fig:expProc} provides an overview of the methodology in this study and the details are described in the following.  

\subsection{Participants}  
In this study, we recruited 74 university students, each of whom received a compensation of \$25 for their participation, involving a duration of approximately one hour. Four participants were excluded from further analysis due to missing data. Data from the remaining 70 participants (mean age = 21.3, SD = 3.0; age range: 18 to 33; 32 female and 38 male participants) were utilized for subsequent analysis. All participants satisfied the requisite criterion of holding a valid driver's license with normal or corrected-to-normal vision.

\begin{figure}[tb!]
\centering
\includegraphics[width=.9\linewidth]{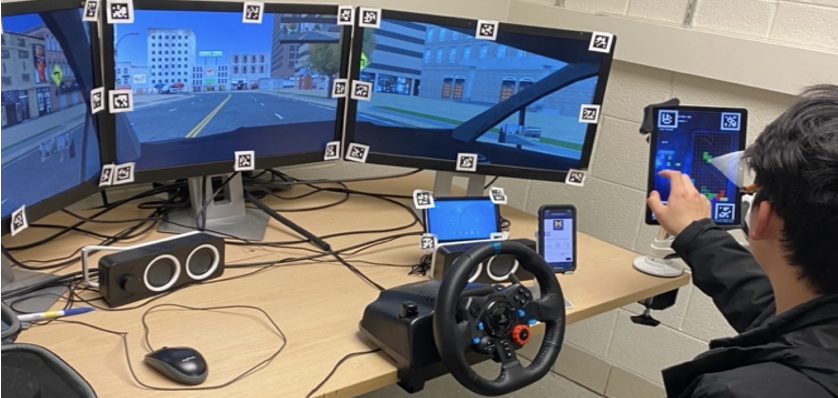}\hfill
\caption{Experiment setup.}
\label{fig:setup}
\end{figure}

\subsection{Apparatus and stimuli}
The study employed a desktop-based driving simulator from Realtime Technologies Inc. (RTI, MI, USA) for data collection, as illustrated in Fig. \ref{fig:setup}. The simulation setup comprised three LCD monitors, integrated with a Logitech driving kit, and two touchscreens (i.e., a tablet and phone), located on the participant's right side, for the non-driving related task (NDRT) and trust rating entry. The NDRT involved a custom-designed Tetris game, implemented using the PyGame library in Python. Participants were required to drag tiles to navigate the game, which they could pause to respond to takeover requests (TORs), and then resume from the same position afterward. Trust was evaluated using a questionnaire created with Qualtrics (Provo, UT, www.qualtrics.com) on a mobile phone. Participants were prompted to rate their trust levels on a scale ranging from 0 to 10, using a single-item question (``How much do you trust the AV?") every 25 seconds, following previous studies \cite{desai2013, ayoub2022real}.

The driving simulation was designed to simulate SAE Level 3 automation. To activate the automated mode, participants were required to depress a red button situated on the steering wheel. Upon initiation of the automated mode, an auditory message, ``Automated mode engaged," would be emitted. In the event of a TOR, participants would be alerted by an auditory warning of ``Takeover" and the automated mode would be promptly disengaged to facilitate the participant's control of the vehicle. In instances where participants failed to resume control of the vehicle within the designated time frame, an auditory emergency stop warning (``Emergency Stop") would be activated to avert any potential collisions.

\subsection{Experimental design}
In this study, we employed a between-subjects design in which participants were assigned to one of the three conditions: a control condition where all eight TORs were valid, a false alarm condition that consisted of four valid TORs and four TORs with false alarms (the 2nd, 3rd, 5th, and 6th), and a miss condition where four of the eight TORs were valid, while the AV missed the other four TORs (the 2nd, 3rd, 5th, and 6th). This approach was adopted to evaluate the impact of different levels of automation performance on trust, as previous studies indicated that both false alarms and misses reduced trust in automated systems \cite{pop2015individual}.

\begin{figure}[tb!]
\centering
\includegraphics[width=.9\linewidth]{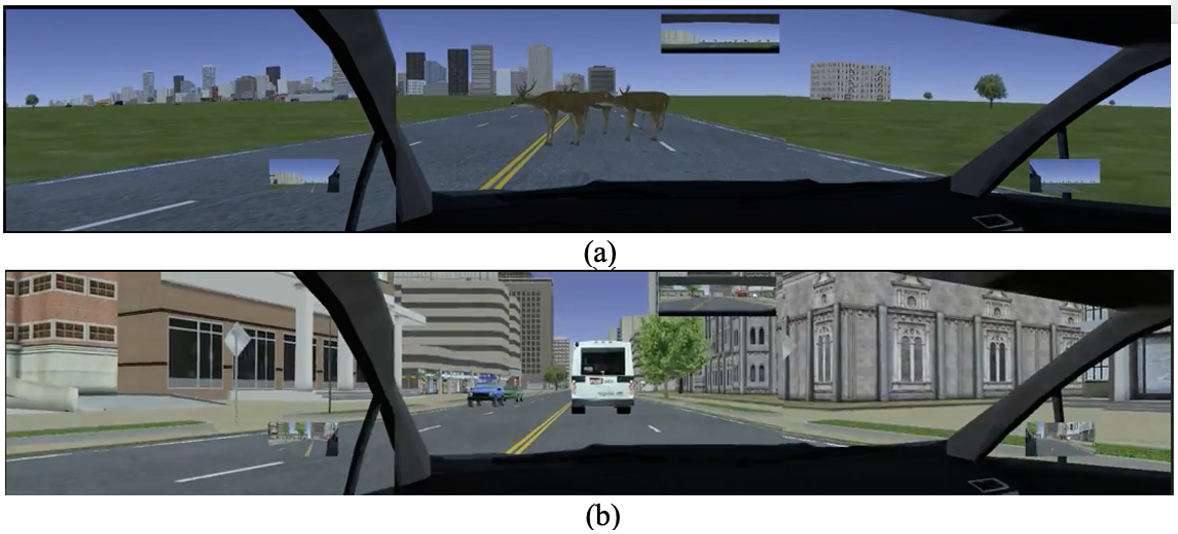}\hfill
\caption{Examples of takeover event in (a) a suburban area with dears ahead (b) an urban area with a bus sudden stop ahead.}
\label{fig:events1}
\end{figure}

During the simulation, standard roadway features were used to signify the occurrence of a TOR event, such as the presence of deer, bicyclists, pedestrians, construction zones, vehicle sudden stops, buses with sudden stops, and police vehicles on the shoulder (see Figures \ref{fig:events1}). The TOR events were equally distributed between rural and urban areas, with four TOR events taking place in each location. The order of the rural and urban scenarios was alternated to minimize potential order effects.

\subsection{Experimental procedure}
Upon arrival, participants were asked to complete a consent form and an online demographic survey. After that, participants received an introduction and watched a short video about the tasks they were required to do. Then, participants completed an online survey that consisted of personality information, propensity to trust AVs, and initial learned trust. The Big Five Model was used to examine different personality traits \cite{zuckerman1993comparison} of the participants using 5-point Likert scales (see Table \ref{tab:personality}). 
The dispositional trust was measured using a six-item trust scale proposed by Merritt et al. \cite{merritt2013trust}.
The statements (see Table \ref{tab:trusts}) were rated with 7-point Likert scales. 
The initial learned trust was assessed with ten items (see Table \ref{tab:trusts}) %\cite{}%
using 7-point Likert scales. 
Subsequently, the participants underwent a training session to familiarize themselves with the driving simulator and experimental protocol. They were required to assess the automated mode functionality of the vehicle and instructed to maintain vigilance and assume control when necessary. Furthermore, they were informed about potential system failure scenarios, including instances of failing to detect obstacles (in the miss condition) and false alarms of TORs (in the FA condition). There were two drives, including urban and suburban with approximately 15 minutes each. The entire experiment was completed within a duration of approximately 60 minutes. To monitor their trust levels dynamically, participants were prompted to respond to a single-item trust rating, ranging from 0 to 10, every 25 second \cite{desai2013impact,ayoub2022real}, with the aim of avoiding any excessive mental or emotional stress \cite{hergeth2016}. Additionally, they were asked to rate their trust levels after completing the driving simulation \cite{desai2013}. Finally, participants rated their anticipated emotional responses to AVs using a 7-point Likert scale that consisted of 19 emotion items, including disdainful, scornful, contemptuous, hostile, resentful, ashamed, humiliated, confident, secure, grateful, happy, respectful, nervous, anxious, confused, afraid, freaked out, lonely, and isolated. These emotions have previously been used to investigate trust in human-machine automation \cite{jensen2020anticipated,avetisian2022anticipated}.

\section{Results}

\subsection{Identifying trust profiles}
We attempted to identify trust profiles using a clustering model based on various measures recorded during the experiment, taking into account factors such as personality, emotions, dispositional, initial learned, and real-time dynamic trust, as well as the performance of the AV (i.e., control, false alarm and misses).  To do this, 48 features from these measures were first normalized between 0 and 1 and then used as input to the K-means clustering model, which was performed in the Azure Machine Learning Studio environment, with the number of cluster chosen as 3 to be the optimal number.  As a result, three clusters were formed and named as follows: 1) \emph{oscillators}, consisting of 23 participants, 2) \emph{believers}, consisting of 31 participants, and 3) \emph{disbelievers}, consisting of 16 participants. %The specifics of each cluster are provided in the subsequent sections.

\begin{figure*}[ht!]
\centering
\includegraphics[width=.6\linewidth]{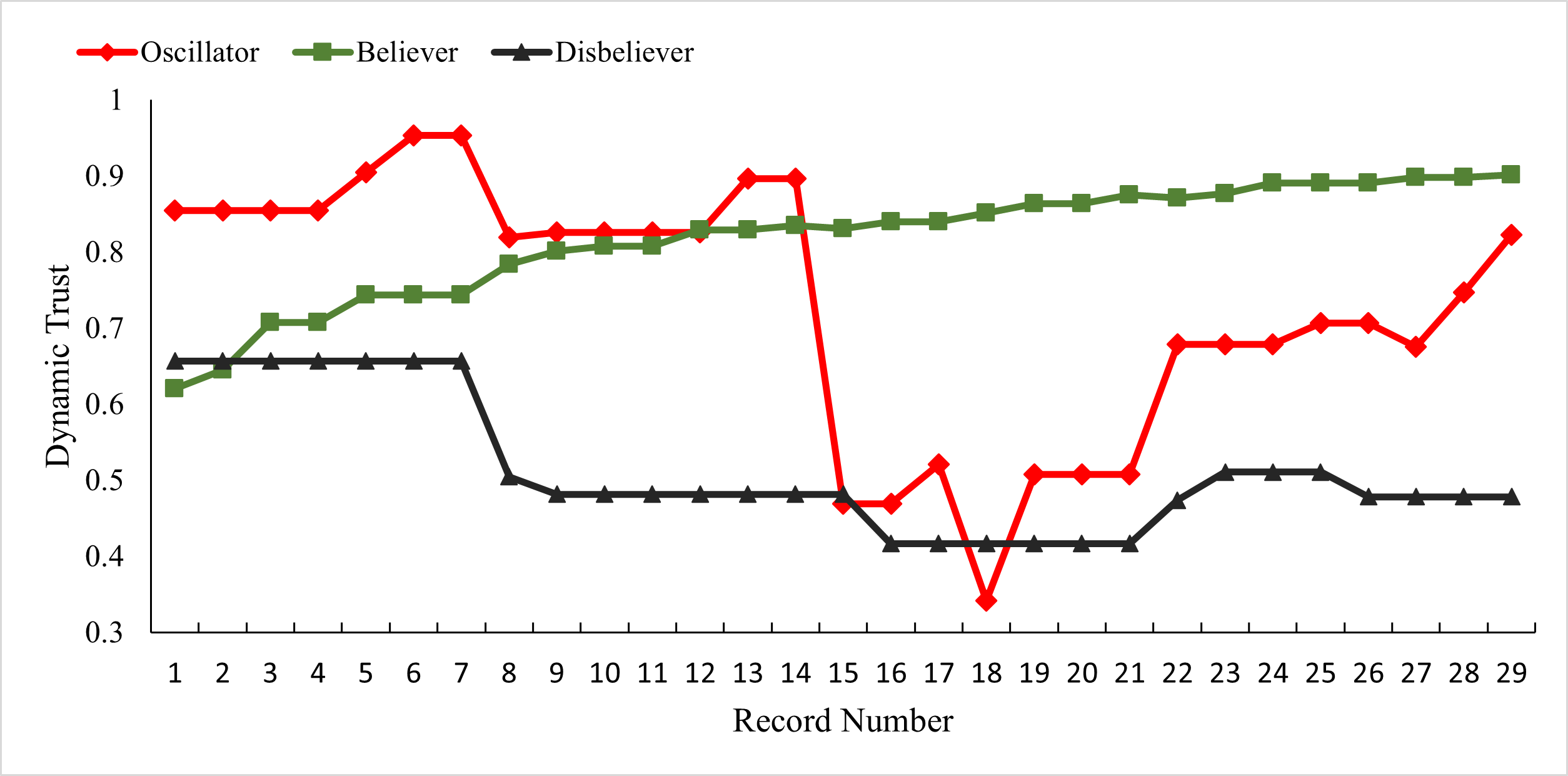}\hfill
\caption{Dynamic trust evaluation of three types of trust profiles during the experiment where the x-axis is the ordinal number of trust ratings in the experiment. Note the number of points of each participant were different and we aggregated them by the order of their self-reported trust to show the dynamics of trust of three profiles.}
\label{fig:personas}
\end{figure*}

Examining the trust profiles, we found that first profile (i.e., \emph{oscillators}) corresponded to drivers who exhibited a moderate level of dynamic trust (Mean = 0.61 and SD = 0.23), but reported high levels of dispositional (Mean = 0.67 and SD = 0.19) and initial learned trust (Mean = 0.65 and SD = 0.22). The individual observations showed that participants’ dynamic trust varied from high to low and vice versa during the experiment. Therefore, this trust profile was referred to as the \emph{oscillators}. Despite the fact that \emph{oscillators} reported the highest levels of dispositional and initial learned trust, their trust level were dynamically evolving based on their most recent experiences resulting in rapid oscillation (see Fig. \ref{fig:personas}).

The second trust profile, referred to as \emph{believers},  consisted of drivers who showed a relatively high level of dynamic trust (Mean = 0.69 and SD = 0.15) during the experiment, but reported moderate levels of dispositional (Mean = 0.53 and SD = 0.22) and initial learned trust (Mean = 0.55 and SD = 0.21). In this profile, the dynamic trust showed an increasing trend during the experiment as participants rated their trust considering their previous knowledge and experience besides the most recent ones happening during the experiment

The last profile (i.e., \emph{disbelievers}) included drivers who exhibited the lowest levels of dynamic trust (Mean = 0.45 and SD = 0.26), as well as low levels of dispositional (Mean = 0.40 and SD = 0.24) and initial learned trust (Mean = 0.42 and SD = 0.24). The individual data observations of \emph{disbelievers} showed that their levels of dynamic trust either decreased or remained low throughout the experiment. Also, drivers in this profile tended to be carping rather than being optimistic about the situation and the AV's overall performance. Therefore, this trust profile was referred to as \emph{disbelievers}.

\subsection{Personality, Emotions, Dispositional and Initial Learned Trust}

\begin{figure}[]
\centering
\includegraphics[width=\columnwidth]{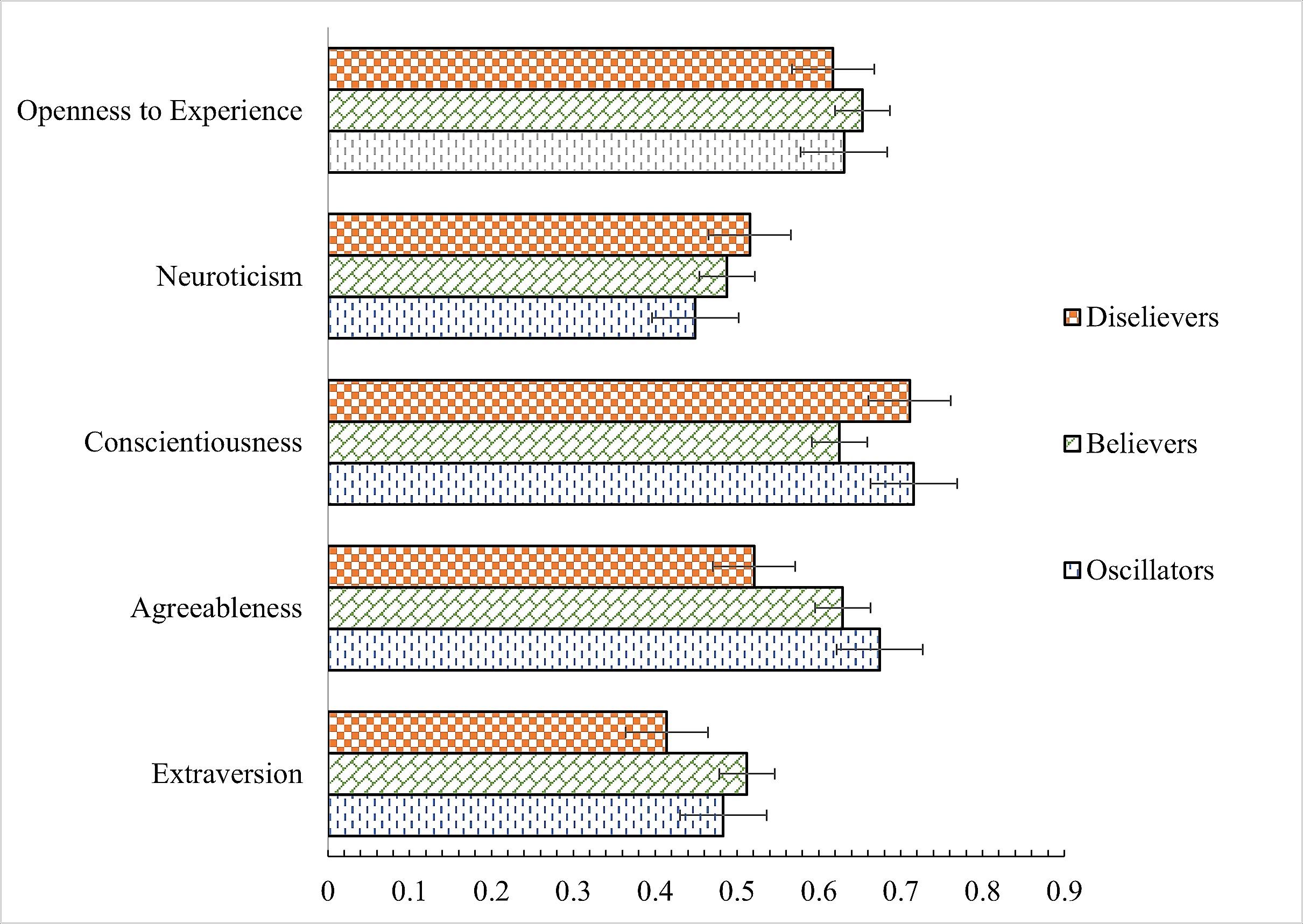}\hfill
\caption{Mean and standard deviation (SD) of Big Five factors (dimensions) of personalities among trust profiles.}
\label{fig:personality}
\end{figure}

\textbf{Personality:} We evaluated the personality of the participants using the Big Five Inventory scale (BFI-10) \cite{rammstedt2007personaliyty}, which measured character traits in five dimensions as shown in Table \ref{tab:personality}: Extraversion (i.e, sociable and reversing reserved), Agreeableness (i.e, trusting and reversing carper), Conscientiousness (i.e, meticulous and reversing lazy), Neuroticism (i.e, nervous and reversing relaxed), and Openness to Experience (i.e, imaginative and reversing artistic). We conducted a one-way ANOVA test to compare these dimensions across the trust profiles and found a significant difference in Agreeableness $(F(2,67)= 2.452, p = 0.033)$. The post-hoc analysis with Tukey-Kramer test showed that \emph{oscillators} had significantly higher levels of Agreeableness compared to \emph{disbelievers} (see Fig. \ref{fig:personality}). In particular, a significant difference was found in the item “I see myself as someone who tends to find fault with others” statement $(p=0.015)$ where \emph{disbelievers} had higher ratings than \emph{believers} and \emph{oscillators} (see Table \ref{tab:personality}). However, there were no significant differences in the other four personality characteristics (i.e., Extraversion, Conscientiousness, Neuroticism, and Openness to Experience). This result was consistent with the SHAP model outcome, as only the ``carper" feature, which represented the above mentioned question in the personality feature set, was selected as an important feature in the model.

% \begin{table}[]
% \caption{Mean, standard deviation(SD) and $p$alue of personality features.}
% \label{tab:personality1}
% \resizebox{\columnwidth}{!}{%
% \begin{tabular}{lllll}
% \hline
% \textbf{Feature}  & \textbf{Cluster 0} & \textbf{Cluster 1} & \textbf{Cluster 2} & \textbf{$p$alue} \\ \hline
% %Reserved & 0.51 (0.27) & 0.49 (0.36) & 0.65 (0.31) & $ >0.05$ \\ 
% %Tend to carp & 0.39 (0.28) & 0.38 (0.28) & 0.62 (0.25) & 0.000 \\ 
% Extraversion &	0.48 (0.17) &	0.51 (0.25) &	0.41 (0.23) &	0.298\\
% Agreeableness ^{\ast} & 0.67 (0.18) &	0.63 (0.25) &	0.52 (0.20) &	0.033\\
% Conscientiousness & 0.72 (0.17) & 0.63 (0.24) & 0.71 (0.19) & 0.254 \\
% Neuroticism & 0.45 (0.21) & 0.49 (0.23) & 0.52 (0.24) & 0.586\\
% Openness to Experience & 0.63 (0.15) & 0.65 (0.20) & 0.62 (0.24) & 0.835\\

% \hline
% \hline
% \end{tabular}%
% }
% \end{table}

\begin{table}[]
\caption{Mean, standard deviation (SD) and $p$-value of individual items in personality. Note: The profiles that share the same letter (i.e., a, b or c) are not statistically different.}
\label{tab:personality}
\resizebox{\columnwidth}{!}{
    \begin{tabular}{lllll}
    \hline    
    \hline
    \textbf{Feature}  & \textbf{\emph{oscillators}} & \textbf{\emph{believers}} & \textbf{\emph{disbelievers}} & \textbf{$p$-value} \\ \hline
    Reserved & 0.51 (0.27) & 0.49 (0.36) & 0.65 (0.31) & 0.085 \\ 
    Trusting & 0.77 (0.21) & 0.65 (0.32) & 0.67 (0.34) & 0.305\\
    Lazy & 0.46 (0.28) & 0.53	(0.30) & 0.47 (0.34) & 0.622\\
    Relaxed & 0.66 (0.21) & 0.60 (0.29) & 0.53 (0.34) & 0.344\\
    Artistic & 0.51 (0.30) & 0.52	(0.33) & 0.48 (0.37) & 0.950\\
    Sociable & 0.74 (0.22) & 0.75	(0.25) & 0.58 (0.36) & 0.101\\
    Carper *** & 0.39 (0.28) $_{a}$ & 0.38 (0.28) $_{ab}$ & 0.62 (0.25) $_{c}$ & 0.000 \\ 
    Meticulous & 0.87 (0.15) & 0.78	(0.25) & 0.89 (0.13) & 0.127\\
    Nervous & 0.57 (0.26) & 0.57 (0.27) & 0.56 (0.28) & 0.991\\
    Imaginative & 0.76 (0.19) & 0.82 (0.22) & 0.72 (0.27) & 0.291\\
    
    \hline

    \end{tabular}
}
\end{table}

\begin{figure}[]
\centering
\includegraphics[width=\linewidth]{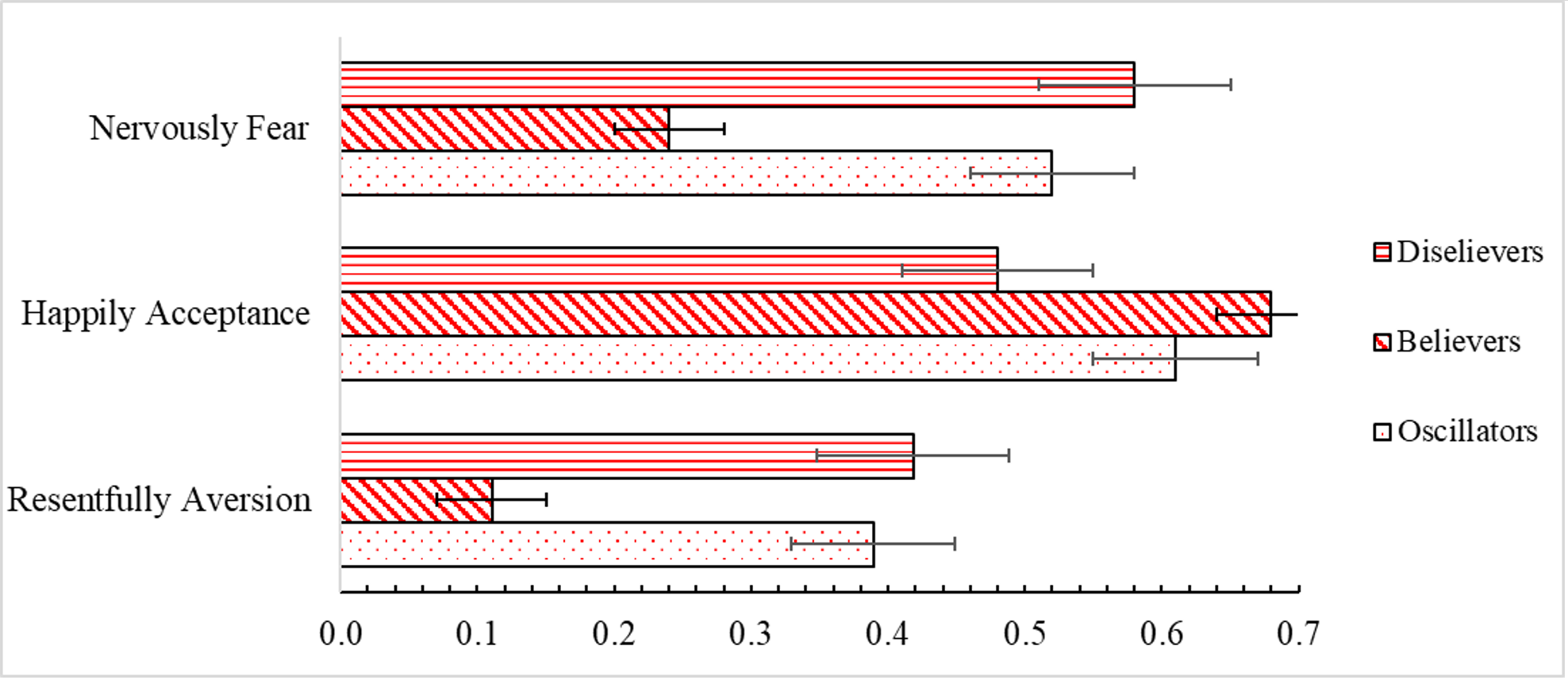}\hfill
\caption{Mean and standard deviation (SD) of three emotion categories among trust profiles.}
\label{fig:grp-emotion}
\end{figure}

\textbf{Emotions:} The results of the one-way ANOVA test indicated that there were significant differences in all the emotions among the three trust profiles, except ``Grateful", as detailed in Table \ref{tab:emotions}. To understand the underlying structure of emotions associated with each trust profile, we conducted an exploratory factor analysis (EFA) and identified three subsets that combined the correlated emotions, i.e \textit{resentfully aversion}: disdainful, scornful, contemptuous, hostile, resentful, ashamed, humiliated, \textit{happily acceptance}: confident, grateful, secure, happy, respectful, not lonely, not isolated, and \textit{nervously fear}: nervous, anxious, confused, afraid, freaked out.
The results showed that \emph{believers} had significantly lower scores for emotions related to resentfully aversion compared to the other trust profiles ($p = 0.000$). They also had significantly higher scores for confidence and security in the happily acceptance category compared to \emph{oscillators}, and higher scores for happiness, gratitude, and respect compared to \emph{disbelievers}. In the nervously fear category, \emph{believers} showed significantly lower scores than the other trust profiles ($ p = 0.000$). Moreover, participants in the \emph{oscillator} profile had significantly higher scores for happiness ($p = 0.016$) compared to \emph{disbelievers} (see Fig. \ref{fig:grp-emotion}). 
%Although the one-way ANOVA test showed that the differences of positive emotions were significant between trust profiles, not all the positive emotions were included for trust profile prediction model, except happy which was highly correlated with other positive emotions. 

\begin{table}[]
\caption{Mean, standard deviation (SD) and $p$-value of emotions. Note: The profiles that share the same letter (i.e., a, b or c) are not statistically different.}
\label{tab:emotions}
\resizebox{\columnwidth}{!}{
\begin{tabular}{lllll}
\hline
\hline
\textbf{Feature}  & \textbf{\emph{oscillators}} & \textbf{\emph{believers}} & \textbf{\emph{disbelievers}} & \textbf{$p$-value} \\ \hline \hline
Lonely & \cellcolor[HTML]{ACDF87} 0.18 (0.22) $_{a}$ & \cellcolor[HTML]{e5f2d9} 0.06 (0.17) $_{ab}$ & \cellcolor[HTML]{76BA1B} 0.44 (0.35) $_{c}$ & 0.000\\
Isolated & \cellcolor[HTML]{ACDF87}  0.20 (0.22) $_{a}$ & \cellcolor[HTML]{e5f2d9} 0.08 (0.17) $_{ab}$ & \cellcolor[HTML]{76BA1B} 0.44 (0.35) $_{c}$ & 0.000\\
Resentful & \cellcolor[HTML]{76BA1B} 0.30 (0.27) $_{a}$ & \cellcolor[HTML]{e5f2d9} 0.10 (0.20) $_{b}$ & \cellcolor[HTML]{76BA1B} 0.34 (0.31) $_{ac}$ & 0.003\\
Humiliated & \cellcolor[HTML]{76BA1B} 0.30 (0.30) $_{a}$ & \cellcolor[HTML]{e5f2d9} 0.06 (0.12) $_{b}$ & \cellcolor[HTML]{76BA1B} 0.36 (0.40) $_{ac}$ & 0.000\\
Ashamed & \cellcolor[HTML]{76BA1B} 0.33 (0.32) $_{a}$ & \cellcolor[HTML]{e5f2d9} 0.06 (0.13) $_{b}$ & \cellcolor[HTML]{76BA1B} 0.39 (0.39) $_{ac}$ & 0.000\\
Hostile & \cellcolor[HTML]{76BA1B} 0.37 (0.27) $_{a}$ & \cellcolor[HTML]{e5f2d9} 0.10 (0.25) $_{b}$ & \cellcolor[HTML]{4C9A2A}  0.56 (0.38) $_{ac}$ & 0.000\\
Disdainful  & \cellcolor[HTML]{76BA1B} 0.41 (0.22) $_{a}$ & \cellcolor[HTML]{e5f2d9} 0.09 (0.18) $_{b}$ & \cellcolor[HTML]{76BA1B} 0.39 (0.28) $_{ac}$ & 0.000\\
Afraid & \cellcolor[HTML]{76BA1B} 0.43 (0.32) $_{a}$ & \cellcolor[HTML]{ACDF87} 0.20 (0.24) $_{b}$ & 
 \cellcolor[HTML]{76BA1B} 0.50 (0.23) $_{ac}$ & 0.000\\
Scornful & \cellcolor[HTML]{76BA1B} 0.45 (0.35) $_{a}$ & \cellcolor[HTML]{ACDF87} 0.13 (0.29) $_{b}$ & \cellcolor[HTML]{76BA1B} 0.38 (0.32) $_{ac}$ & 0.001\\
Freaked out & \cellcolor[HTML]{76BA1B} 0.45 (0.32) $_{a}$ & \cellcolor[HTML]{e5f2d9} 0.06 (0.14) $_{b}$ & \cellcolor[HTML]{76BA1B} 0.40 (0.22) $_{ac}$ & 0.000\\
Grateful  & \cellcolor[HTML]{4C9A2A} 0.54 (0.23) & \cellcolor[HTML]{26580F} \textcolor{white}{0.62 (0.21)} & \cellcolor[HTML]{76BA1B} 0.48 (0.20) & 0.080\\
Nervous & \cellcolor[HTML]{4C9A2A} 0.54 (0.25) $_{a}$ & \cellcolor[HTML]{76BA1B} 0.33 (0.26) $_{b}$ & \cellcolor[HTML]{26580F} \textcolor{white}{0.64 (0.20) $_{ac}$} & 0.000\\
Confused & \cellcolor[HTML]{4C9A2A} 0.56 (0.24) $_{a}$ & \cellcolor[HTML]{76BA1B} 0.24 (0.26) $_{b}$ & 
 \cellcolor[HTML]{26580F} \textcolor{white}{0.68 (0.18) $_{ac}$} & 0.000\\
Contemptuous & \cellcolor[HTML]{4C9A2A} 0.57 (0.35) $_{a}$ & \cellcolor[HTML]{76BA1B} 0.23 (0.33) $_{b}$ & \cellcolor[HTML]{4C9A2A} 0.52 (0.27) $_{ac}$ & 0.001\\
Secure & \cellcolor[HTML]{4C9A2A} 0.57 (0.23) $_{a}$ & \cellcolor[HTML]{26580F} \textcolor{white}{0.75 (0.14) $_{b}$} & \cellcolor[HTML]{76BA1B} 0.46 (0.25) $_{ac}$ & 0.000\\
Confident & \cellcolor[HTML]{4C9A2A}  0.59 (0.25) $_{a}$ & \cellcolor[HTML]{26580F} \textcolor{white}{0.72 (0.12) $_{b}$} & \cellcolor[HTML]{4C9A2A} 0.55 (0.15) $_{ac}$ & 0.003\\
Happy & \cellcolor[HTML]{4C9A2A} 0.59 (0.27) $_{a}$ & \cellcolor[HTML]{26580F} \textcolor{white}{0.63 (0.20) $_{ab}$} & \cellcolor[HTML]{76BA1B} 0.34 (0.20) $_{c}$ & 0.000\\
Respectful & \cellcolor[HTML]{26580F} \textcolor{white}{0.61 (0.27) $_{a}$} & \cellcolor[HTML]{26580F} \textcolor{white}{0.72 (0.21) $_{ab}$} & \cellcolor[HTML]{76BA1B} 0.45 (0.19) $_{ac}$ & 0.001\\
Anxious & \cellcolor[HTML]{26580F} \textcolor{white}{0.66 (0.27) $_{a}$} & \cellcolor[HTML]{76BA1B} 0.38 (0.27) $_{b}$ & \cellcolor[HTML]{26580F} \textcolor{white}{0.70 (0.24) $_{ac}$} & 0.000\\
\hline
\end{tabular}%
}
\end{table}

\textbf{Dispositional, Initial Learned, and Dynamic Trust:} With regard to dispositional and learned trust, the results of one-way ANOVA showed that there was a significant difference between three trust profiles (see Table \ref{tab:trusts}) for all the factors measuring dispositional and initial learned trusts ($ p <0.05$), except one from initial learned trust that measured trust in delegating control task to AV when driving was boring ($p=0.08$).
As for dynamic trust, the results showed that \emph{disbelievers} had significantly different trust levels in the urban and suburban areas. Specifically, in the urban area, \emph{disbelievers} had a lower trust level compared to the other trust profiles ($p = 0.000$), while in the suburban areas, the difference was significant compared to \emph{believers} ($p = 0.017$). %Moreover,we found differences in the change patterns in trust during the experiment. Further details about these findings are described in the next section. 

%  \begin{figure}[]
% \centering
% \includegraphics[width=1\columnwidth]{Figures/dispositional-heatmap.png}\hfill
% \caption{Heatmap showing the distribution of dispositional trust across different clusters.}
% \label{fig:dispTrust}
% \end{figure}

%  \begin{figure}[]
% \centering
% \includegraphics[width=.9\columnwidth]{Figures/learned-heatmap.jpg}\hfill
% \caption{Heatmap showing the distribution of initial learned trust across different clusters.}
% \label{fig:learnedTrust}
% \end{figure}

\begin{table*}[]
\caption{Mean, standard deviation (SD) and $p$-value of dispositional and initial learned trust. Note: The profiles that share the same letter (i.e., a, b or c) are not statistically different.}
\label{tab:trusts}
\resizebox{\linewidth}{!}{
\begin{tabular}{llllll}
\hline
\hline
&\textbf{Feature}  & \textbf{\emph{oscillators}} & \textbf{\emph{believers}} & \textbf{\emph{disbelievers}} & \textbf{$p$-value} \\ \hline
 \multirow{6}{3mm}{\begin{turn}{90}Dispositional \end{turn}} 
 & I usually trust self-driving vehicles until there is a reason not to & \cellcolor[HTML]{26580F} \textcolor{white}{0.81 (0.16) $_{a}$} & \cellcolor[HTML]{26580F} \textcolor{white}{0.67 (0.20) $_{b}$} & \cellcolor[HTML]{ACDF87} 0.45 (0.26) $_{c}$ & 0.000\\
& For the most part, I distrust self-driving vehicles &\cellcolor[HTML]{ACDF87} 0.24 (0.21) $_{a}$ & \cellcolor[HTML]{ACDF87} 0.39 (0.21) $_{b}$ & \cellcolor[HTML]{4C9A2A} 0.57 (0.25) $_{c}$ & 0.000\\
& In general, I would rely on a self-driving vehicle to assist me & \cellcolor[HTML]{26580F} \textcolor{white}{ 0.78 (0.18) $_{a}$} & \cellcolor[HTML]{26580F} \textcolor{white}{0.61 (0.19) $_{b}$} & \cellcolor[HTML]{ACDF87} 0.41 (0.27) $_{c}$ & 0.000\\
& My tendency to trust self-driving vehicles is high & \cellcolor[HTML]{26580F} \textcolor{white}{ 0.78 (0.19) $_{a}$} & \cellcolor[HTML]{4C9A2A} 0.53 (0.21) $_{b}$ & \cellcolor[HTML]{ACDF87} 0.35 (0.24) $_{c}$ & 0.000\\
& It is easy for me to trust self-driving vehicles to do their job & \cellcolor[HTML]{26580F} \textcolor{white}{0.78 (0.16) $_{a}$} & \cellcolor[HTML]{4C9A2A} 0.56 (0.21) $_{b}$ & \cellcolor[HTML]{ACDF87} 0.42 (0.25) $_{bc}$ & 0.000\\
& I am likely to trust self-driving vehicles even when I have little knowledge about it & \cellcolor[HTML]{26580F} \textcolor{white}{0.66 (0.27) $_{a}$} & \cellcolor[HTML]{ACDF87} 0.42 (0.28) $_{b}$ & \cellcolor[HTML]{ACDF87} 0.23 (0.19) $_{bc}$ & 0.000\\
\hline
\multirow{10}{3mm}{\begin{turn}{90}Initial learned\end{turn}} & I would feel safe in a self-driving vehicle & \cellcolor[HTML]{26580F} \textcolor{white}{0.80 (0.11) $_{a}$} & \cellcolor[HTML]{26580F} \textcolor{white}{0.65 (0.16) $_{b}$} & \cellcolor[HTML]{ACDF87} 0.46 (0.22) $_{c}$ & 0.000\\
&The self-driving vehicle system provides me with more safety & \cellcolor[HTML]{26580F} \textcolor{white}{0.70 (0.25) $_{a}$} & \cellcolor[HTML]{ACDF87} 0.45 (0.20) $_{b}$ & \cellcolor[HTML]{ACDF87} 0.39 (0.22) $_{bc}$ & 0.000\\
& I would rather keep manual control of my vehicle than delegate it to the self-driving vehicle system on every occasion & \cellcolor[HTML]{ACDF87} 0.39 (0.30) $_{a}$ & \cellcolor[HTML]{4C9A2A} 0.57 (0.29) $_{ab}$ & \cellcolor[HTML]{26580F} \textcolor{white}{0.76 (0.18) $_{bc}$} & 0.000\\
& I would trust the self-driving vehicle system decisions & \cellcolor[HTML]{26580F} \textcolor{white}{0.76 (0.14) $_{a}$} & \cellcolor[HTML]{26580F} \textcolor{white}{0.65 (0.13) $_{ab}$} & \cellcolor[HTML]{ACDF87} 0.43 (0.25) $_{c}$ & 0.000\\
& I would trust the self-driving vehicle system capacities to manage complex driving situations & \cellcolor[HTML]{26580F} \textcolor{white}{0.64 (0.24) $_{a}$} & \cellcolor[HTML]{ACDF87} 0.42 (0.21) $_{b}$ & \cellcolor[HTML]{ACDF87} 0.33 (0.28) $_{bc}$ & 0.000\\
& If the weather conditions were bad, I would delegate the driving task to the self-driving vehicle system & \cellcolor[HTML]{ACDF87} 0.45 (0.30) $_{a}$ & \cellcolor[HTML]{ACDF87} 0.35 (0.25) $_{ab}$ & \cellcolor[HTML]{ACDF87} 0.22 (0.19) $_{bc}$ & 0.030\\
& Rather than monitoring the driving environment, I could focus on other activities confidently & \cellcolor[HTML]{26580F} \textcolor{white}{0.61 (0.24) $_{a}$} & \cellcolor[HTML]{ACDF87} 0.48 (0.29) $_{ab}$ & \cellcolor[HTML]{ACDF87} 0.26 (0.24) $_{c}$ & 0.001\\
& If driving was boring for me, I would rather delegate it to the self-driving vehicle system than do it myself & \cellcolor[HTML]{26580F} \textcolor{white}{0.70 (0.27)} & \cellcolor[HTML]{26580F} \textcolor{white}{0.67 (0.21)} & \cellcolor[HTML]{4C9A2A} 0.52 (0.28) & 0.081\\
& I would delegate the driving to the self-driving vehicle system if I was tired & \cellcolor[HTML]{26580F} \textcolor{white}{0.79 (0.24) $_{a}$} & \cellcolor[HTML]{26580F} \textcolor{white}{0.74 (0.20) $_{ab}$} & \cellcolor[HTML]{4C9A2A} 0.59 (0.29) $_{bc}$ & 0.042\\
& I would trust the self-driving vehicle & \cellcolor[HTML]{26580F} \textcolor{white}{0.76 (0.15) $_{a}$} & \cellcolor[HTML]{4C9A2A} 0.55 (0.20) $_{b}$ & \cellcolor[HTML]{ACDF87} 0.30 (0.26) $_{c}$ & 0.000\\

\hline
\end{tabular}%
}
\end{table*}

\subsection{Validating Trust Profile Models}
In order to validate the trust profile models, we used the cluster membership as the ground truth and tested the accuracy of the clustering model by training supervised machine learning models with all the 48 features. We used an Azure automated ML experiment job to identify the best machine learning model in Azure automated ML experiment by fine-tuning the hyperparameters. The multinomial logistic regression model was found to perform the best across a large number of models, including XGBoost, LightGBM, Random Forest, etc. The results showed that the model was able to predict the trust profile with F1-score = 0.90. %(see Table \ref{tab:model} for the model performance summary). 

\begin{figure}[tb!]
\centering
\includegraphics[width=.9\linewidth]{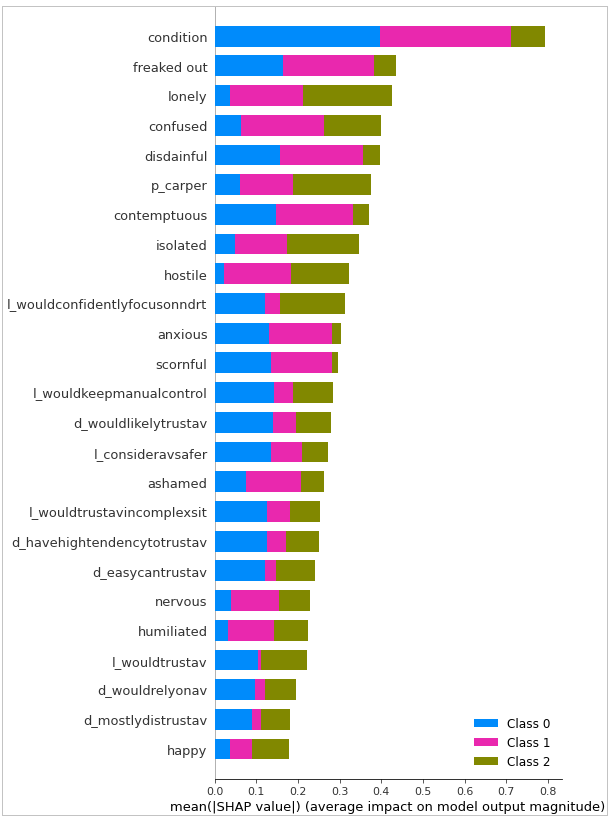}\hfill
\caption{Summary of SHAP values where the features are ordered from the highest to the lowest effect on the prediction. In the figure, the prefix ``p{\_}" in features indicates personality-related items, ``d{\_}" indicates dispositional trust items, and ``l{\_}" indicates learned trust items.}
\label{fig:shap}
\vspace{-10pt}
\end{figure}

\begin{figure*}[ht!]
\centering
\includegraphics[width=0.7\linewidth]{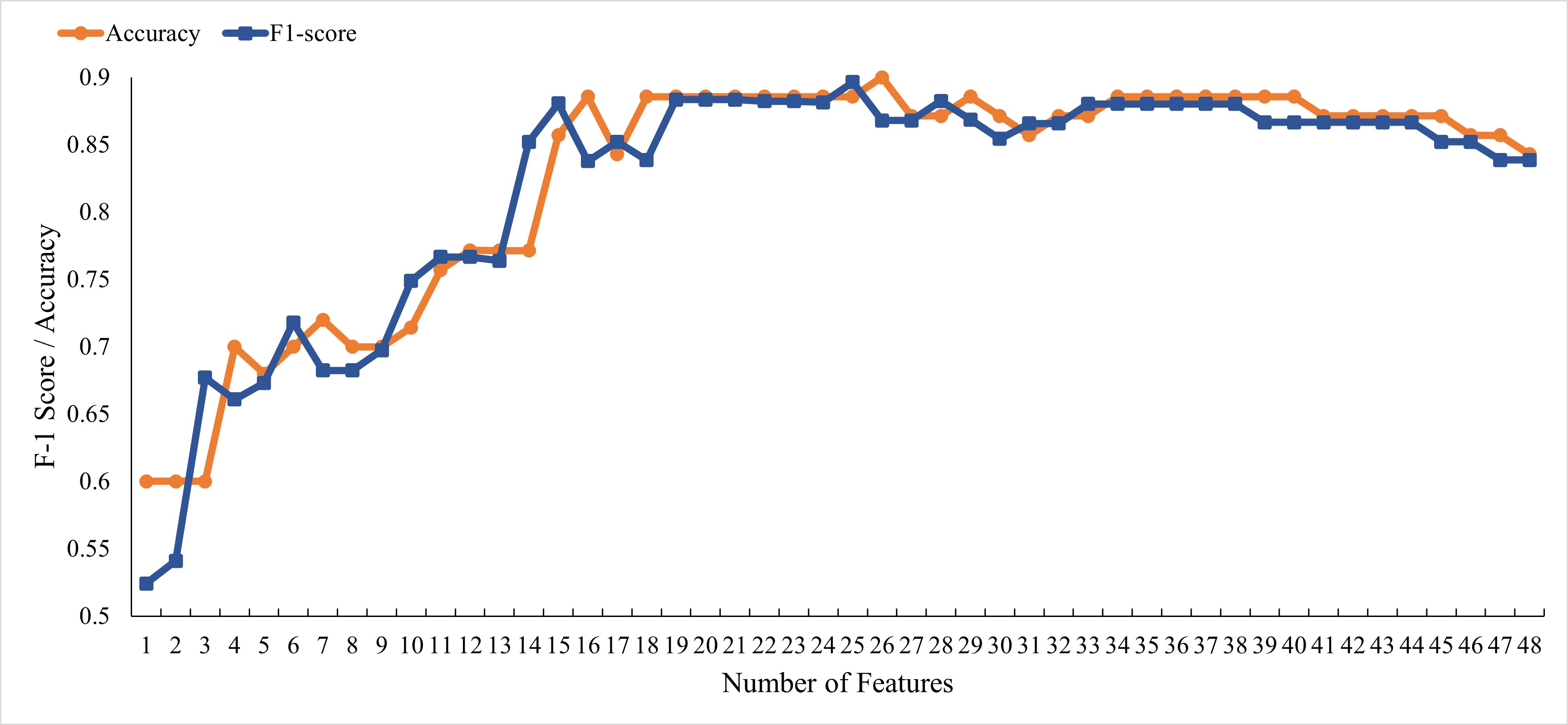}\hfill
\caption{The impact of individual features on overall performance of the model. During each iteration, a single feature was added to the model based on SHAP importance ranking.}
\label{fig:perf}
\end{figure*}

We also used SHapley Additive exPlanations (SHAP) explainer to interpret the model and understand the importance of individual features in the model's decision-making process. Fig. \ref{fig:shap} illustrates the feature importance rankings, from the most influential to the least influential on the prediction. The features of condition (miss, false alarm, or control), emotions of freaked out and confused were found to be the most significant ones in predicting the trust profiles, regardless of whether they had a positive or negative effect. The color coding indicates the feature importance in distinguishing between different trust profiles. We used a feature selection method to improve the model's performance while preserving the patterns and relationships in the data, as suggested by Ayoub et al. \cite{ayoub2022cause} and \cite{ayoub2022predicting}. We added one feature at a time based on the feature importance ranking from the SHAP explainer and validated the results with stratified 5-fold cross validation. As shown in Fig. \ref{fig:perf}, the model reached the highest performance (F1-score = 0.90) using the top 25 features, i.e. condition, \textit{emotions}: freaked out, lonely, confused, disdainful, contemptuous, isolated, hostile, ashamed, anxious, scornful, nervous, humiliated, happy, \textit{personality}: carper,  \textit{initial learned trust}: I could confidently focus on NDRT, I would keep manual control, I would consider AV more safe than manuals, I would trust in complex situations, I would trust AV, \textit{dispositional trust}: I likely to trust AV without knowledge, I have high tendency to trust AV, I easily can trust AV,  I would rely on an AV, I mostly distrust AV). 
% \textit{emotions}: i.e, Condition, Freaked out, Lonely, Confused, Disdainful,  \textit{personality}: p{\_}to{\_}carp, Contemptuous, Isolated, Hostile, \textit{initial learned}: l{\_}focus{\_}on{\_}ndrt, Anxious, Scornful, l{\_}keeps{\_}man{\_}control, d{\_}likely{\_}trust, l{\_}more{\_}safety, Ashamed, l{\_}trust{\_}to{\_}manage{\_}complex{\_}situations, \textit{dispositional} d{\_}high{\_}tendency, d{\_}easy{\_}trust, Nervous, Humiliated, l{\_}trust{\_}toAV, d{\_}will{\_}rely{\_}on{\_}AV, d{\_}mostly{\_}distrust, Happy)
Overall, across all the 48 features, the F1-score ranged from 0.52 to 0.90, and the accuracy ranged from 0.60 to 0.90.

\begin{figure}[tb!]
\centering
\includegraphics[width=1\linewidth]{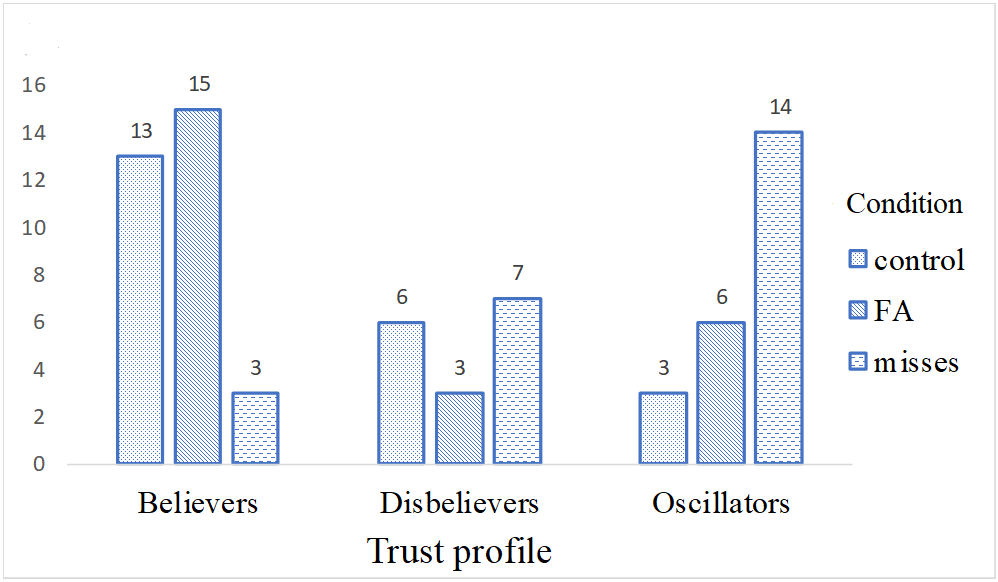}\hfill
\caption{The distribution of the clusters across the three conditions. }
\label{fig:condition}
\vspace{-10pt}
\end{figure}

Even though the experiment condition was found to be the most influential feature in predicting these three trust profiles, it was not sufficient to determine the identified trust profiles alone. As shown in Fig. \ref{fig:condition}, there was no one-to-one mapping relationships between the participants in three conditions, including control, FA, and misses and those formed the three identified trust profiles. Also shown in Fig. \ref{fig:perf}, the condition alone only had F1-score = 0.52 and accuracy = 0.60 in predicting trust profiles. Other factors, such as emotions, personality, initial learned trust, and dispositional trust, also played important roles in determining these trust profiles.

\section{Discussion}

\subsection{Identifying Trust Profiles}
In this study,  we aimed to identify trust profiles of drivers in conditionally automated driving using a multidimensional dataset that included drivers’ personality, dispositional trust, initial learned trust, dynamic trust levels, and emotions. First, we performed a clustering method to group the drivers exhibiting similar behaviors and unveil the relationships between the investigated features. 
We identified three types of trust profiles, namely \emph{oscillators}, \emph{believers}, and \emph{disbelievers}. Then, we used a multinomial logistic regression model to validate the trust profile clustering model and the results showed that the model could predict the driver trust profile with an F1-score of 0.90. The important features selected by SHAP explainer indicated that the formation of dynamic trust in AV was highly associated with system performance (e.g., false alarms, misses) emotions, initial learned trust, dispositional trust, and personality, etc. The three conditions in the experiment did not result in three distinct trust profiles. Instead other factors also played important roles. 

The trust profile of \emph{oscillators} was characterized by a moderate level of dynamic trust, which fluctuated based on the recent experiences. This profile corresponded to drivers who were generally optimistic about the AV's performance and in some situations were willing to take risks instead of taking over the vehicle. Nevertheless, the emotional state of \emph{oscillators} was significantly impacted when the AV's performance included errors, indicating that the trust oscillation had an impact on the intensity of particular emotions \cite{avetisian2022anticipated}. 
%These oscillations were also related to the particular takeover situations, as in some cases participants experienced trust decrease when they felt that time was not enough to take over the AV safely. 

The trust profile of \emph{believers} was characterized by a relatively high level of dynamic trust, which increased over time during the experiment. This profile corresponded to drivers who were confident in AV's abilities to control the vehicle and were willing to delegate the driving task to the system. From the emotion point of view, \emph{believers} had a more positive attitude towards AV performance and were less prone to fear or disappointment when encountering failures. Specifically, they felt a greater sense of safety and were less vulnerable to false alarms, even if the takeover requests were unfounded, in contrast to \emph{disbelievers} who regarded such incidents as severe system malfunctions.

The \emph{disbelievers}, who were characterized by the lowest level of dynamic trust, tended to be pessimistic and displayed a low level of trust even when the AV performed properly. In addition to trust reports, the significantly high ratings of nervousness and fear (see Table \ref{tab:emotions}), as well as for the manual control and monitoring attributes in the learned trust (see Table \ref{tab:trusts}) during the experiment verified that \emph{disbelievers} were not willing to delegate the driving task to the vehicle or be engaged in NDRTs. This perspective might be due to the limited understanding of the underlying processes of AV systems or prior negative experiences with automated systems.

Regarding personality differences, the trust profiles showed similarities except for the trait of agreeableness towards the system. \emph{Disbelievers} had a significant disagreement with the system and were confident that they could perform better than the AV, despite the AV performance.

The identified trust profiles exhibited patterns with the trust bases outlined by previous studies stating that trust was always based on at least one of three characteristics of an automation system, i.e. purpose, performance, and process \cite{mayer1995, lee2004trust}. Specifically, those who were categorized as \emph{believers}, the purpose was the dominant characteristic for trust and if the automation failed, the affected trust dimension was the performance-related trust \cite{lee1992trust}. While \emph{disbelievers} trusted the AV system based on the process, meaning that they did not trust the AV until they understood the decision making process. \emph{Oscillators} trusted the system based on its performance based on the AV's capabilities to perform specific tasks in certain situations, and the trust change depended on their comprehension of how the situation impacted on AV's capabilities.

%According to Stokes et al. \cite{stokes2010mood}, individuals who experience positive emotions (e.g., \emph{believers}) tended to have higher levels of trust compared to those experiencing negative emotions. Additionally, positive emotions impacted one's tendency to take risks \cite{zhou2014prospect}, which potentially could lead to overtrust in AVs. While there were no significant differences in emotions between \emph{oscillators} and \emph{disbelievers}, except for feelings of lonely, isolated and happy \ref{tab:emotions}, Fig. \ref{fig:shap} shows that when combined with history-based trust features (i.e., dispositional and initial learned trust items), the effect of some emotion notably changed with regard to the dynamic trust, which described distinct behavioral patterns. In terms of personality features, our findings did not reveal any significant correlations between trust profiles and personality. Thus, further studies are required to explore and understand the potential influence of personality factors on trust dynamics.

\subsection{Implications}

The identification of three distinct trust profiles, namely \emph{believers}, \emph{disbelievers}, and \emph{oscillators}, has significant implications for the design of AVs.

For \emph{believers}, who exhibit high levels of trust in AVs, the design should aim to provide features that increase their confidence and comfort levels, such as clear and intuitive interfaces, well-defined communication protocols, and robust safety features that effectively communicate the system's status and capabilities.  They also tend to relate to the purpose of trust \cite{mayer1995, lee2004trust} and believe that automation could help improve efficiency, reduce errors, and enhance safety by delegating their tasks to machines and freeing up their own cognitive and physical resources to focus on other tasks. This could be not limited to automated driving but in various domains, such as healthcare, aviation, and manufacturing. 

However, believers exhibited significantly more positive emotions, which might impact their tendency to take risks \cite{zhou2014prospect} and potentially lead to overtrust in AVs. To address this issue, the design should incorporate clear and transparent communication that explains the system's capabilities and limitations by clearly showing the limitations of the system \cite{jiang2020effects}, and include educational materials to help them understand how the technology works. %For example, the system should be designed to clearly communicate when the system would request the driver to take over control. %Moreover, designing systems that require the driver to periodically take control of the vehicle would help prevent overtrust by reminding the driver that they are ultimately responsible for the safety of the vehicle \cite{lee2004trust}.

\emph{Disbelievers}, who exhibit low levels of trust in AVs, present a unique challenge in the design process for them to adopt such technologies.  First, the design should incorporate effective training and education to build trust and confidence over time. More important, they tend to relate more to the process of the automation \cite{mayer1995, lee2004trust}, which might help improve their trust levels in automated driving through transparency and predictability with continuous feedback. If the vehicle is designed so that they can easily understand what the system is doing and why it is doing it through clear and intuitive communication as feedback, this would be helpful \cite{avetisyan2022investigating}. For example, by including explaining why and what will information with speech and augmented reality during the takeover process, participants reported to be easy to use and accept SAE Level 3 vehicles \cite{du2021designing}. We should also consider predictability of the system so that the system's behavior aligns with user expectations and how well it performs in different scenarios. For example, when participants received explanations about the vehicle's behavior ahead of the time and their possible projection in the future, they had better situation awareness of the driving scenarios and trust in automated driving \cite{avetisyan2022investigating}. These two strategies can help drivers to understand the system's behavior and performance, which in turn can enhance their learned trust and confidence in the system. 

Finally, for \emph{oscillators}, whose trust in AVs fluctuates based on their most recent experience, the design should aim to deliver a consistently positive experience. 
This can be achieved by providing reliable and consistent performance across a variety of situations, such as inclement weather, heavy traffic, and challenging road conditions. In this aspect, adaptive automation might help \emph{oscillators} to calibrate their trust by adapting the level of automation based on the drivers' current trust level in AV and the system's past performance \cite{moray2000adaptive}.
For instance, if the driver is confident and happy in the AV, the system can perform in higher levels of autonomy and require less intervention from the driver. On the other hand, if the driver is less confident in the AV resulted from previous negative experiences, the system can provide feedback about system status and require drivers intervention in order to increase the feelings of confidence, control, and safety. Therefore, integrating the adaptive automation into the AV system, designers would help \emph{oscillators} to calibrate their trust in the automation system, improving overall system performance and safety \cite{moray2000adaptive, kaber2005adaptive}.

\subsection{Limitations}
The present study has several limitations that require further investigation in future research. Firstly, the study was conducted using a low-fidelity experimental setup utilizing a desktop driving simulator. The sample size of 70 participants was relatively small and we only considered a limited number of factors to identify trust profiles. Moreover, the study sample was composed primarily of university students, which resulted in a homogeneous sample, regarding age, education, driving experience, and knowledge about AVs. To overcome these limitations, future studies should be conducted in higher fidelity experimental settings and with a larger and more diverse sample size. More external factors should be included, such as time constraints, perceived risks, complexity of NDRTs, and situation awareness \cite{hoff2015, manchon2021manualtoAD, avetisyan2022investigating}. 

Second, the assessment of trust was predominantly based on takeover scenarios in which participants' dynamic trust in control, misses, and FA conditions was measured. The study only covered a limited number of potential takeover issues. Therefore, future research should explore trust in other scenarios, such as continuous driving performance in object detection, different driving styles, and route selection recommendations. Overcoming these limitations would allow for a better generalization of the trust profiles to other human-machine interaction scenarios (e.g., \cite{bhat_jessie2022clusteringTrust}).

Finally, we only included a limited number of factors that formed the trust profiles. While these factors provided valuable insights into the dynamics of trust, they may not fully capture the complexity and diversity of trust. It is important to acknowledge that trust is a multifaceted construct influenced by various individual and contextual factors, which might not be fully accounted for in this study. Therefore, the generalizability of the study findings to other contexts and populations might be limited and future studies should include a wider range of trust indicators to enhance the external validity of the results.

\section{Conclusions}

The purpose of our research was to examine the trust profiles in AVs and determine the underlying behavioral patterns of drivers that could be useful for designing profiles-based systems. To accomplish this, we collected multidimensional data and clustered them into three trust profiles: \emph{believers}, \emph{oscillators}, and \emph{disbelievers}. We used these profiles to build a logistic regression model that could predict the trust profile with accuracy of 0.89 and F1-score of 0.90. Additionally, we used SHAP explainer to identify the most significant factors in the dataset that influenced the creation of trust profiles. Furthermore, we investigated the dynamic trust patterns among these profiles, as well as the associated initial learned trust, dispositional trust, emotional and personality characteristics, based on which we discussed how to develop a system that can adjust AV's behavior based on the driver's trust profiles, and eventually promote their acceptance and adoption of such technologies.

\section{Acknowledgement}
This research was supported by National Science Foundation.

\bibliography{main}
\begin{IEEEbiography}[{\includegraphics[width=1in,height=1.55in,clip,keepaspectratio]{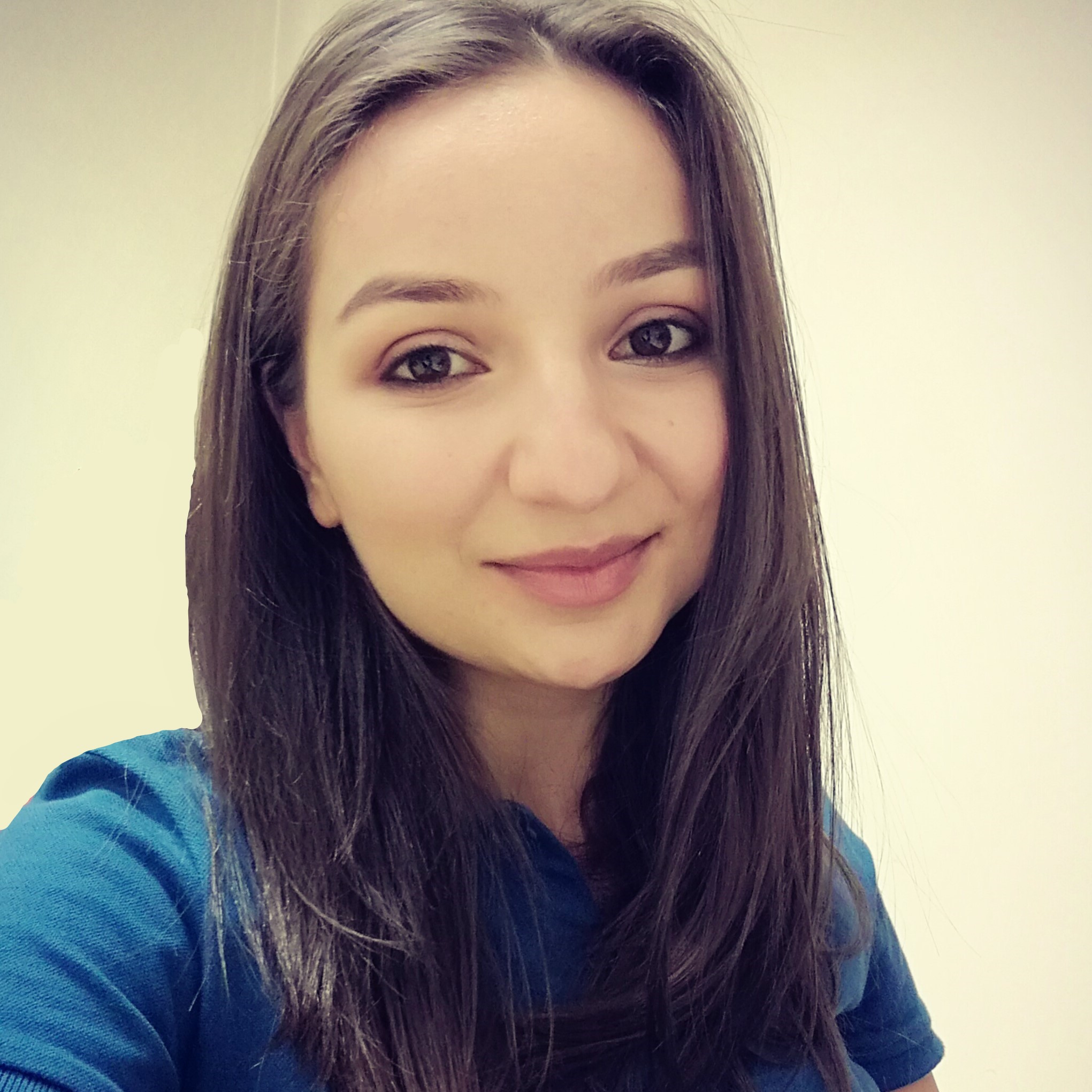}}]{Lilit Avetisyan received her B.E. degree in 2017 and MS degree in 2019 in Information Security from the National Polytechnic University of Armenia. She is currently pursuing her Ph.D. degree in Industrial and Systems Engineering at the University of Michigan, Dearborn. Her main research interests include human-computer interaction, explainable artificial intelligence and situation awareness.}
% Her research aims to improve human performance and safety by applying human factors and data analytics techniques to the analysis, design, and evaluation of the technologies. 
\vspace{-20pt}
\end{IEEEbiography}

\begin{IEEEbiography}[{\includegraphics[width=1in,height=1.55in,clip,keepaspectratio]{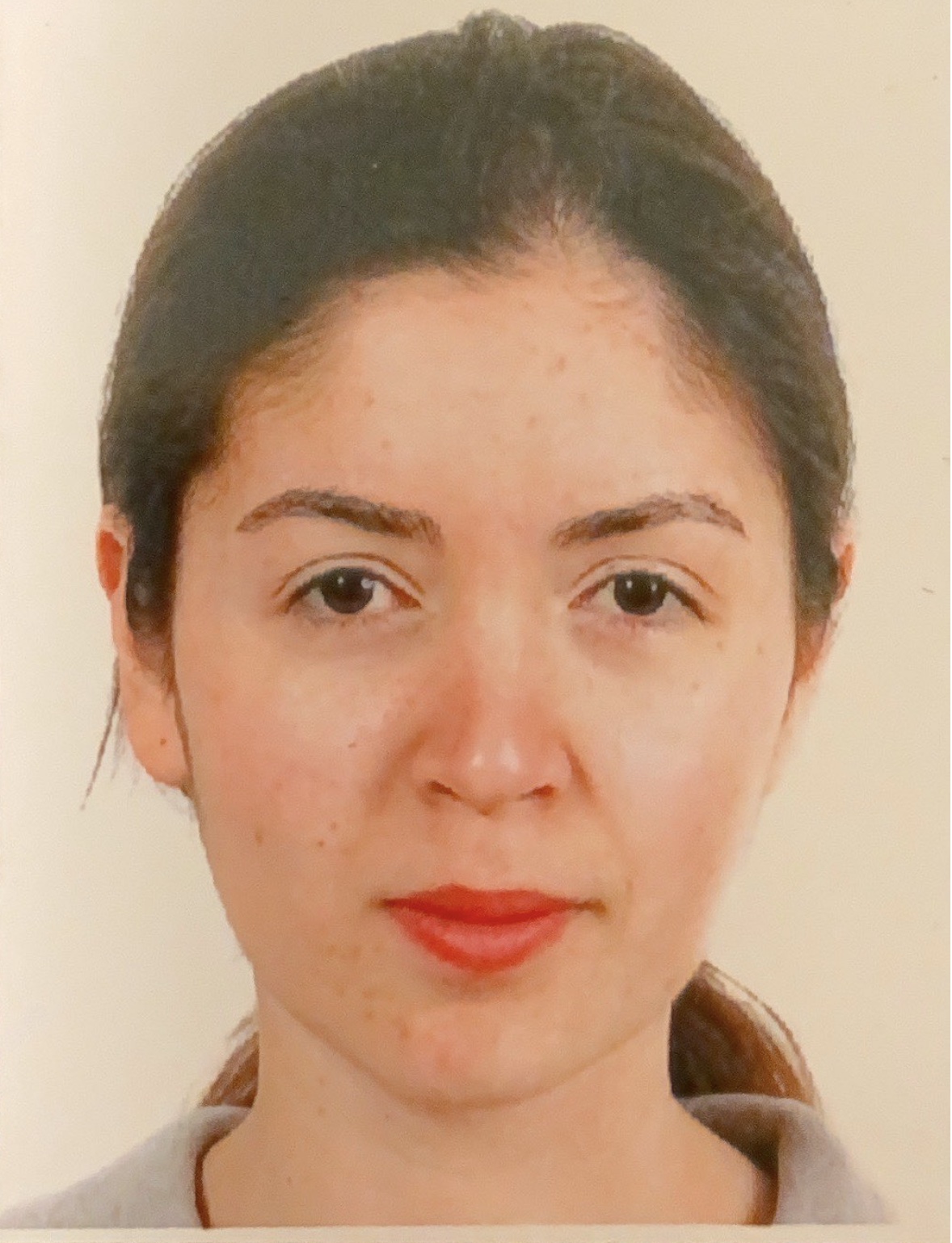}}]{Jackie Ayoub received her B.E. degree in mechanical engineering from Notre Dame University, Lebanon, in 2016 and her master degree in Industrial and Systems Engineering from University of Michigan, Dearborn, in 2017. She received her Ph.D. in Industrial and Systems Engineering in the University of Michigan, Dearborn, in 2022. She is currently a data scientist and human factor engineer at Honda Research Institute, Detroit. Her main research interests include human-computer interaction, human factors and ergonomics, and sentiment analysis.}
\vspace{-20pt}
\end{IEEEbiography}
%\vskip 0pt plus -1fil

% Na Du has been publishing in human factors and transportation research venues.
%\vskip 0pt plus -1fil
\begin{IEEEbiography}[{\includegraphics[width=1in,height=1.55in,clip,keepaspectratio]{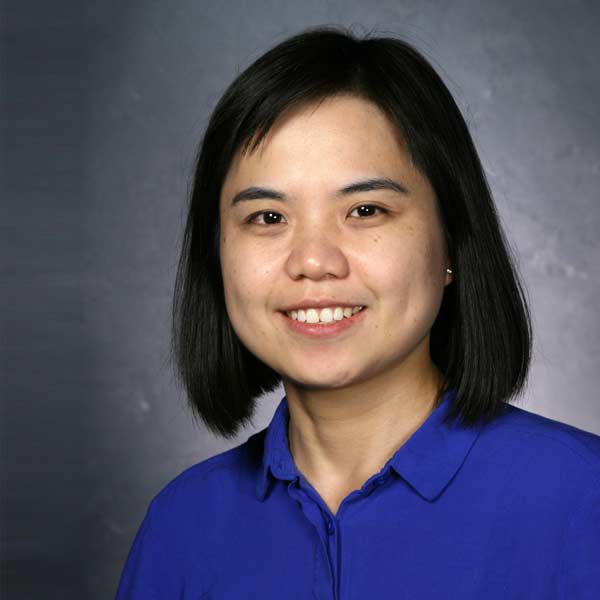}}]{X. Jessie Yang is an Assistant Professor in the Department of Industrial and Operations Engineering, University of Michigan, Ann Arbor. She earned a PhD in Mechanical and Aerospace Engineering (Human Factors) from Nanyang Technological University, Singapore. Dr. Yang’s research include human-autonomy interaction, human factors in high-risk industries and user experience design.}
\vspace{-20pt}
\end{IEEEbiography}

%\vskip 0pt plus -1fil
\begin{IEEEbiography}[{\includegraphics[width=1in,height=1.55in,clip,keepaspectratio]{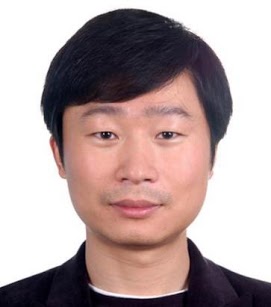}}]{Dr Feng Zhou received the Ph.D. degree in Human Factors Engineering from Nanyang Technological University, Singapore, in 2011 and Ph.D. degree in Mechanical Engineering from Gatech Tech in 2014. He was a Research Scientist at MediaScience, Austin TX, from 2015 to 2017. He is currently an Assistant Professor with the Department of Industrial and Manufacturing Systems Engineering, University of Michigan, Dearborn. His main research interests include human factors, human-computer interaction, engineering design, and human-centered design.}
\end{IEEEbiography}

%\vfill


\end{document}